\documentclass[journal=acscii,manuscript=article]{achemso}

\usepackage[version=3]{mhchem} 
\usepackage{pdfpages}
\usepackage{tabularx}
\usepackage{subcaption}
\usepackage{booktabs}
\usepackage{graphicx}

\newcommand{\siref}[1]{({SI Appendix#1})}
\newcommand{\fref}[1]{Fig. \ref{#1}}
\newcommand{\tref}[1]{Table \ref{#1}}
\newcommand{\labelphantom}[1]{%
  \parbox{0pt}{\phantomsubcaption\label{#1}}%
}
\graphicspath{{paper/}}
\author{Mgcini Keith Phuthi}
\affiliation
{Department of Mechanical Engineering, Carnegie Mellon University, Pittsburgh, PA, USA}
\author{Archie Mingze Yao}
\affiliation{Department of Mechanical Engineering, Carnegie Mellon University, Pittsburgh, PA, USA}
\author{Simon Batzner}
\author{Albert Musaelian}
\author{Boris Kozinsky}
\affiliation{School of Engineering and Applied Science, Harvard University, Cambridge, MA, USA}
\author{Ekin Dogus Cubuk}
\affiliation{Google Research, Brain Team}
\author{Venkatasubramanian Viswanathan}
\affiliation{Department of Mechanical Engineering, Carnegie Mellon University, Pittsburgh, PA, USA}
\altaffiliation{Corresponding Author}
\email{venkvis@cmu.edu}

\title{Accurate Surface and Finite Temperature Bulk Properties of Lithium Metal at Large Scales using Machine Learning Interaction Potentials}

\abbreviations{MLIP,ML,LMSSB}
\keywords{Batteries, Lithium, Interatomic Potentials, Machine Learning, Molecular Dynamics}

\begin{document}






    

\begin{abstract}
The properties of lithium metal are key parameters in the design of lithium ion and lithium metal batteries. They are difficult to probe experimentally due to the high reactivity and low melting point of lithium as well as the microscopic scales at which lithium exists in batteries where it is found to have enhanced strength, with implications for dendrite suppression strategies [1]. Computationally, there is a lack of empirical potentials that are consistently quantitatively accurate across all properties and ab-initio calculations are too costly. In this work, we train Machine Learning Interaction Potentials (MLIPs) on Density Functional Theory (DFT) data to state-of-the-art accuracy in reproducing experimental and ab-initio results across a wide range of simulations at large length and time scales. We accurately predict thermodynamic properties, phonon spectra, temperature dependence of elastic constants and various surface properties inaccessible using DFT. We establish that there exists a Bell-Evans-Polanyi relation correlating the self-adsorption energy and the minimum surface diffusion barrier for high Miller index facets.
\end{abstract}

\section{Introduction}

Lithium metal batteries provide a promising pathway to achieving high capacity energy storage devices. However, realizing practical lithium metal batteries has been limited by morphological instabilities, primarily related to dendrite formation and thus safety issues \cite{wang_lithium_2020, zhu_design_2020, lin_reviving_2017}. A number of approaches have been proposed to address the issue of dendrite formation. 
One approach is suppressing instability through the introduction of a solid electrolyte in contact with lithium metal.  Monroe and Newman proposed the use of a solid polymer electrolyte with a shear modulus larger than that of lithium \cite{monroe_impact_2005}, which was extended by Ahmad and Viswanathan showing that an alternate approach could be to use a lithium-dense solid electrolyte whose modulus is smaller than that of lithium \cite{ahmad_stability_2017}.  
A second approach increases morphological stability through rapid surface diffusion, quantified by surface diffusion barriers \cite{jackle_self-diffusion_2018, gaissmaier_first_2019}.  All of these approaches critically hinge upon accurate determination of the properties of lithium metal.  
The first approach requires knowing the room temperature mechanical properties of lithium, while the second approach requires a detailed understanding of surface diffusion barriers across high Miller index facets formed during morpohological instability.  Determining these properties experimentally is challenging due to the high reactivity of lithium \cite{li_atomic_2017} thus computational methods provide a more practical approach.

In principle, many material properties can be calculated from first principles using high fidelity, atomistic methods such as Ab-Initio Molecular Dynamics (AIMD). In practice however, the computational cost of these methods is often prohibitive due to poor scaling with the number of electrons ($\sim$O$(N^3)$) and the long simulation time per timestep \cite{kresse_efficiency_1996}. As an alternative, approximate empirical potentials have been commonly used for Molecular Dynamics (MD) simulations to achieve necessary time and length scales to collect good statistics but often at the cost of considerable loss of accuracy leading to only qualitative or often wrong results for some properties. In the last decade, a number of Machine Learning Interaction Potentials (MLIPs) have been developed which demonstrate remarkable accuracy in reproducing ab-initio results compared to empirical potentials if trained with a sufficiently sampled dataset \cite{behler_machine_2021, batzner_e3-equivariant_2022, schutt_schnet_2018}.

MLIPs have been used to study supercritical phenomena in hydrogen \cite{cheng_evidence_2020}, defects in various metals \cite{freitas_machine-learning_2022}, have been benchmarked for transition metals \cite{owen_complexity_2023} and many other examples. For lithium, the SNAP potential \cite{zuo_performance_2020} has been developed for purposes of benchmarking MLIPs and is therefore limited in its applicability as we show in this work. Jiao et al. generated a Deep Potential \cite{zhang_deep_2018} and simulated the self-deposition of Li and the different morphologies that could arise in deposition processes \cite{jiao_self-healing_2022}. Their potential however was not accurate in predicting stresses and elastic constants, which we improve upon in our own Deep Potential. 

In this work, we generate data to train two general purpose MLIPs for pure lithium metal based on NequIP \cite{batzner_e3-equivariant_2022} and Deep Potential. The MLIPs, particularly NequIP, reproduce DFT and experimental results remarkably well over a wide range of structures including, bulk, surfaces, defects and liquids, all in one potential, consistently outperforming empirical potentials and existing MLIPs. We therefore more accurately calculate elastic and surface properties important to the design of lithium metal batteries and discuss the implications of our results.

\section*{Results}
In this section, we first demonstrate the accuracy of the trained MLIPs by comparing the predictions directly to results from DFT and other potentials in the literature. Second, we present results from a number of simulations that are typically too expensive or otherwise impossible to do with DFT and compare them to experiment. Finally we predict key properties in the design of lithium metal batteries using simulations that have, up to this point, been inaccessible using DFT using the NequIP model architecture.

In addition to DFT, where possible, we compare our results with the popular MEAM empirical potential developed by Kim et al. \cite{kim_atomistic_2012} and SNAP MLIP by Zuo et al. \cite{zuo_performance_2020}. The MEAM potential has been used to predict, for example, thermal behavior in lithium metal electrodes \cite{luo_thermal_2022} and the lifetime of glassy lithium nuclei under fast charging conditions \cite{wang_glassy_2020}. SNAP is a MLIP that was designed for benchmarking a number of MLIP architectures and not necessarily built for production simulations. It is worth noting that the differences in predictions with our MLIPs can be attributed not only to the quality of the fit but also to the difference in the datasets used to fit the potentials and therefore the potentials need to be compared with experiment to be assessed rigorously. We perform many of these assessments in the following sections.

We show results for two NequIP potentials trained with different levels of precision. NequIP32 and NequIP64 are trained to single (float32) and double (float64) precision respectively as Batatia et al. found that float32 can be insufficiently precise to represent total energy differences \cite{batatia_design_2022}. In our experiments, we found that the lower precision of NequIP32 led to a more coarse discretization of the potential energy surface resulting in errors of $\sim$10meV for a modest number of $\sim$100 atoms when calculating energy differences. The NequIP32 potential however uses less memory and is faster hence it would be advantageous to use a lower precision, particularly for MD simulations which depend on forces and are not affected by small errors in energy and do not have changing numbers of atoms and we have found to be stable. Properties like the vacancy formation energies however are affected by this precision hence we show both results. We found that the numerical instabilities were improved by performing operations in the last layers of the model architecture in double precision, however we did not implement this solution for the potentials in this work. We present results for both NequIP32 and NequIP64 to highlight any differences. All MD simulations however were performed at single precision to save computational effort. We also show results for our developed Deep Potential (DP). All calculations for DP were performed at double precision.

\subsection{DFT Benchmarks}

\begin{table*}
\centering
\caption{Various properties of BCC lithium predicted using the MLIPs and compared to DFT results in the literature, DFT in this work and the existing MEAM and SNAP potentials. Percentage errors relative to the DFT prediction are shown in square brackets. Except for the Anisotropy, all the errors are within 5\% for the NequIP64 potential. NequIP32 predicts has a large vacancy formation energy error due to the lower precision used to calculate small energy differences. SNAP and MEAM produce different results as they were trained on different data and parameterized differently than in this work.}
\resizebox{\columnwidth}{!}{\begin{tabular}{lccccccc}
Property &  DFT          & DFT        & DeepMD & NequIP32 & NequIP64 & SNAP & MEAM \\
         &  (other work) & (this work)&        &          &          &      &      \\
\midrule
Energy RMSE (meV/atom) & - & - & 3 & 1 & 1 & - & -\\
Force RMSE (meV/\AA) & - & - & 20 & 12 & 12 & - & - \\
Stress RMSE (GPa) & - & - & 0.22 & 0.06 & 0.06 & - & - \\
Lattice Constant (\AA) & 3.427 \cite{zuo_performance_2020} & 3.434 & 3.434 [0.0]& 3.431 [-0.1] & 3.429 [-0.1] & 3.494 [1.7] & 3.506 [2.1] \\
$E_v$ (eV/atom) & 0.62 \cite{zuo_performance_2020} & 0.525  & 0.518 [-1.4] & 0.567 [7.9] & 0.520 [-0.9] & 0.486 [-7.4] & 0.378 [-27.9] \\
Bulk Modulus (GPa) & 14 \cite{zuo_performance_2020}& 13.7 & 13.7 [0.0] & 14.0 [2.2] & 14.0 [2.3] & 10.5 [-23.7] & 12.9 [-5.7] \\
$C_{11}$ (GPa) & 15 \cite{zuo_performance_2020} & 14.8 & 14.2 [-4.3] & 14.9 [0.4] & 14.7 [-0.4] & 18.4 [24.0] & 17.9 [21.2] \\
$C_{12}$ (GPa) & 13 \cite{zuo_performance_2020} & 13.1 & 13.5 [2.4] & 13.6 [3.2] & 13.6 [3.8] & 6.5 [-50.5] & 10.4 [-20.9] \\
$C_{44}$ (GPa) & 11 \cite{zuo_performance_2020} & 10.4 & 13.2 [26.5] & 10.9 [4.7] & 10.9 [4.7] & 10.0 [-3.7] & 12.7 [22.3] \\
Anisotropy & 11 \cite{zuo_performance_2020} & 12.6 & 37.2 [196.1] & 16.8 [33.3] & 19.6 [56.1] & 1.7 [-86.5] & 3.4 [-73.1] \\
(100) Surface Energy (eV/\AA$^2$) & 0.029 \cite{tran_surface_2016} & 0.029 & 0.029 [0.5] & 0.029 [-0.8] & 0.029 [-0.7] & 0.027 [-7.1] & 0.024 [-15.9] \\
(110) Surface Energy (eV/\AA$^2$) & 0.031 \cite{tran_surface_2016} & 0.031 & 0.031 [0.0] & 0.031 [-0.6] & 0.031 [-0.1] & 0.028 [-9.4] & 0.024 [-21.9] \\
(111) Surface Energy (eV/\AA$^2$) & 0.034 \cite{tran_surface_2016} & 0.033 & 0.034 [3.0] & 0.033 [0.9] & 0.033 [0.2] & 0.030 [-8.6] & 0.028 [-14.6] \\
\bottomrule
\end{tabular}}\label{tab:benchmarks}
\end{table*}

Predictions for a number of different properties accessible using standard DFT methods are shown in \tref{tab:benchmarks} to benchmark the MLIPs. Details on how the calculations were performed are given in the Supporting Information \siref{}. 

The conventional metric to assess the accuracy of an MLIP is the Root Mean Square Error (RMSE) on the prediction of the label $y$ which can be energy, forces and stresses over a test set. The RMSE is defined as $$ \text{RMSE} = \sqrt{\sum\limits^N_{i=1}\frac{(\hat{y}_i-y_i)^2}{N}} $$ where $y_i$ is the ground truth label and $\hat{y}_i$ is the predicted label of sample $i$ in a test set with $N$ samples. The RMSEs shown in Table \ref{tab:benchmarks} demonstrate very good accuracy on the order of $\sim$1meV/atom for energies, comparable to the fluctuations in the energy with converged k points and typically chosen DFT convergence criteria \cite{sholl_density_2009}. The forces and stresses are also well converged to $\sim$10meV/\AA{} and $\sim$1meV/\AA$^3$ respectively such that an atomic/cell parameter displacement of $\sim 0.01$\AA{} gives an energy difference of $\sim$1meV/atom. It is worth noting that both NequIP potentials perform an order of magnitude better in stress error (0.06eV/\AA$^3$) compared to DP (0.22eV/\AA$^3$) and are slightly better than DP in most other metrics. We also note that we trained the NequIP architecture on the data that was used to train the SNAP potential and got better test errors on the SNAP test set with little hyperparameter tuning. NequIP RMSEs were 1.3meV/atom and 15mev/\AA{} vs quoted SNAP RMSEs of 1.4meV/atom and and 40meV/\AA{} for SNAP for energy and force RMSE respectively \cite{zuo_performance_2020}. We thereby expect that the NequIP architecture is generally more accurate for the same dataset, consistent with previous results \cite{batzner_e3-equivariant_2022}.

The RMSE is a good metric for benchmarking MLIP architectures against each other but does not imply an accurate potential in predicting experimental properties since it strongly depends on an inevitably biased test set.  We therefore calculate a number of other properties to benchmark the MLIPs which reflect the quality of the distribution of data in the training set as well as the appropriateness of the choice of ground-truth data, particularly the choice of DFT exchange-correlation functional. \tref{tab:benchmarks} shows how all the properties calculated here are in excellent agreement with most being within 5\% of the DFT predictions for the DP and NequIP potentials. An exception is the anisotropy which is derived from the elastic constants as $2C_{44}/(C_{11}-C_{12})$ and hence propagates error from the elastic constants.

\begin{figure*}[!tbh]
    \centering
    \labelphantom{fig:phonons}
    \labelphantom{fig:surface_energies}
    \labelphantom{fig:wulff}
    \includegraphics[width=.8\textwidth]{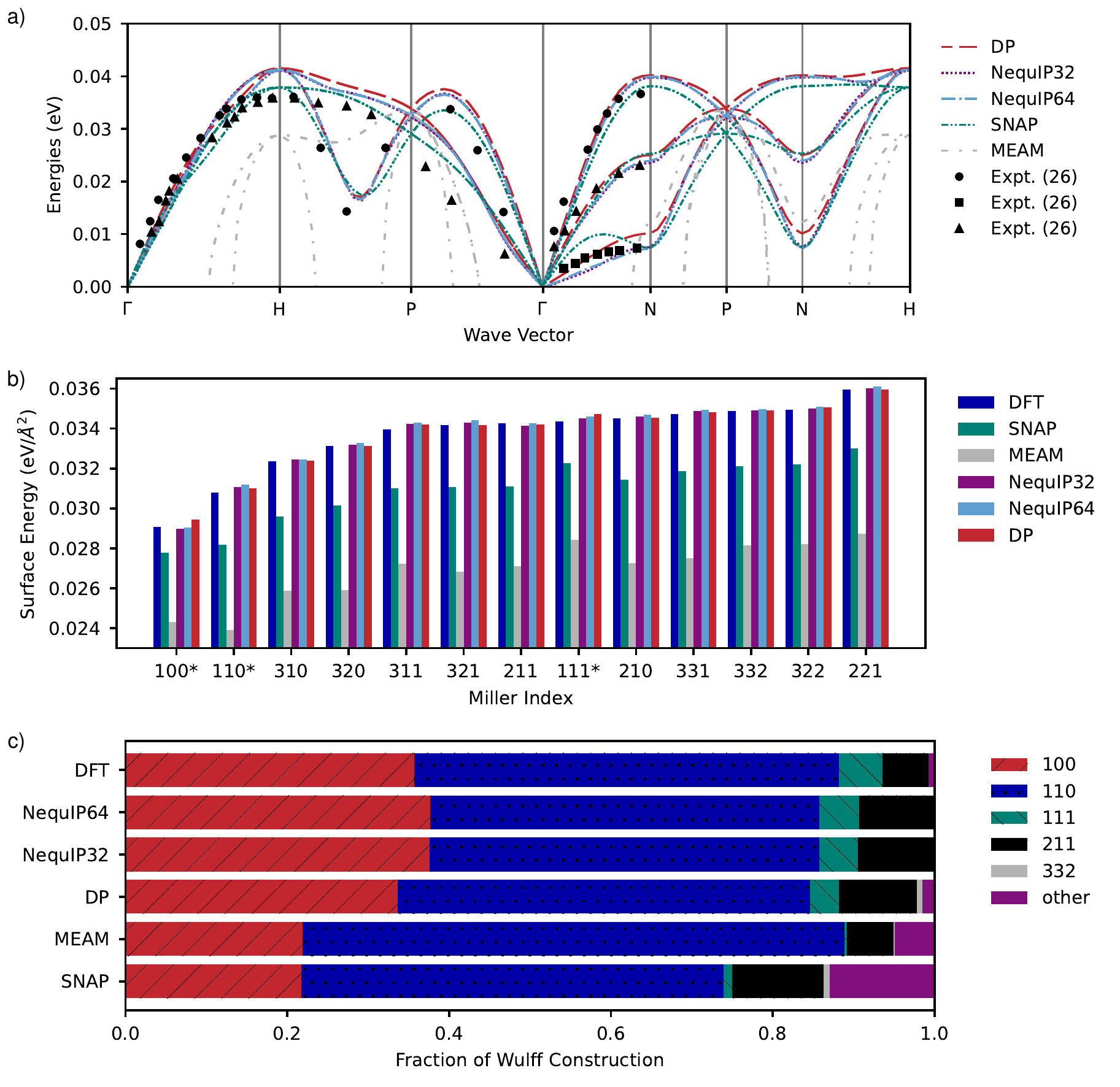}
    \caption{Various properties of BCC lithium calculated and compared with DFT results. a) Phonon spectrum calculated with the various potentials, NequIP and DP from this work as well as SNAP perform well but MEAM completely fails to predict a reasonable spectrum. b) Surface energies for various facets of BCC lithium. NequIP and DP potentials reproduce DFT results very well while MEAM and SNAP have large errors relative to the DFT prediction. c) The Wulff construction for BCC lithium showing that BCC lithium in vacuum is dominated by the (100), (110), (111) and (211) facets according to DFT. NequIP and DP potentials in this work reproduce that result but SNAP and MEAM predict a significant contribution from other facets and much less from (111).}
    \label{fig:surfaces}
\end{figure*}

\subsection{Phonon Spectrum}
We calculate the phonon spectrum of BCC lithium using the \textit{phonons} package as implemented in the Atomic Simulation Environment (ASE) in \fref{fig:phonons}. The finite displacement method was used with a displacement of 0.01\AA{} for a supercell of size 5x5x5. Note that while this supercell is larger than the cutoff in the MLIPs of 6\AA, the NequIP potentials can still predict slight changes in force due to their message passing architecture with multiple layers, resulting in a much larger effective cutoff radius capable of capturing some long range effects \cite{musaelian_learning_2022}.

Calculated phonon band energies are usually higher than the experimental values because they are calculated at 0K whereas the experimental values are at 300K. Beg et al. found that the phonon band energies in BCC lithium decrease with increasing temperature due to changes in the volume of the equilibrium unit cell and can decrease by even up to 10\% for some parts of the Brillouin zone \cite{beg_temperature_1976}. The temperature effects can be taken into account by sampling the dynamical matrix at finite temperature but this is beyond the scope of this work. The structural features of the bands however are clearly well captured by NequIP, DP and SNAP which produces slightly different results, including the broadening/splitting of the phonons between $\Gamma$-H, typical of anisotropic cubic materials like lithium \cite{beg_temperature_1976}. MEAM however completely fails to capture the phonon spectrum as it was not specifically trained on phonon data due to having a short cutoff that does not capture long range interactions important at small wave vectors.  The Modified Analytic Embedde Atom Method (MAEAM) has been found to predict a reasonable spectrum \cite{xie_atomistic_2011}.

\subsection{Surface Energies and Wulff Construction}
As shown in \fref{fig:surface_energies}, in addition to bulk properties, the NequIP and DP potentials accurately describe the surface energies of lithium. Surface and adsorption energies are often quoted up to $\sim$10meV/\AA$^2$ precision due to finite size effects and propagation of errors. The NequIP and DP potentials are well within 1meV/\AA$^2$ error for all the miller indices despite being only explicitly trained on the $(100)$, $(110)$ and $(111)$ planes as starting seeds. This demonstrates the excellent generalization to higher miller index surfaces which we take advantage of in the calculation of surface properties in Section \ref{sec:adsorption}. The predicted surface energies for low miller indices listed in \tref{tab:benchmarks} agree very well with both our DFT and DFT results in other works whereas the MEAM and SNAP predictions give different values and slightly different rankings for the surface energies as a function of miller index suggesting they are less reliable for surface properties.

We also calculate the Wulff construction which determines the equilibrium shape of a droplet or crystal suspended in a medium which in this case is vacuum \cite{balluffi_kinetics_2005}. Depending on the surface energies and geometry, the shape of the droplet determines what fraction of the total area of the droplet is contributed by each facet. This allows us to estimate the importance of considering a particular facet in simulations at equilibrium. In the presence of an electrolyte, the Wulff construction might change due to the sensitivity to the small energy differences between higher miller indices and in practice the surrounding medium \cite{tran_surface_2016}. We follow the procedure used by Tran et al. \cite{tran_surface_2016} implemented in the Python Materials Genomics package (pymatgen) to estimate the area fractions for miller indices up to a value of 3 in \fref{fig:wulff}. The NequIP results agree well with the DFT prediction that (110) and (100) planes followed by (111) and (211) to a lesser extent dominate the Wulff construction. DP has slightly different results with (211) more dominant over (111) while MEAM and SNAP have much more significant differences as expected due to the poor prediciton of surface energies. The domination of (110) in the Wulff construction despite (100) having the lowest surface energy is consistent with experimental results \cite{li_atomic_2017}.
 
The excellent reproduction of DFT calculated properties gives confidence that the MLIPs, particularly NequIP are truly reproducing the DFT result for surface and bulk properties with small errors at the level of typical DFT precision. We therefore assume subsequent errors are from the quality of the dataset, most likely the choice of exchange-correlation functional when comparing with experiment. In the rest of the work, we demonstrate the superior accuracy in reproducing experimental results of the NequIP potentials over a range of properties that are difficult to predict using DFT.

\subsection{Temperature Dependence of Elastic Properties of Lithium}

\begin{figure*}[!tbh]
\centering
\labelphantom{fig:lattice_constant}
\labelphantom{fig:c11}
\labelphantom{fig:c12}
\labelphantom{fig:c44}
\labelphantom{fig:anisotropy}
\labelphantom{fig:kvrh}
\labelphantom{fig:gvrh}
\labelphantom{fig:youngsmodulus}
\labelphantom{fig:poisson}
\includegraphics[width=0.8\linewidth]{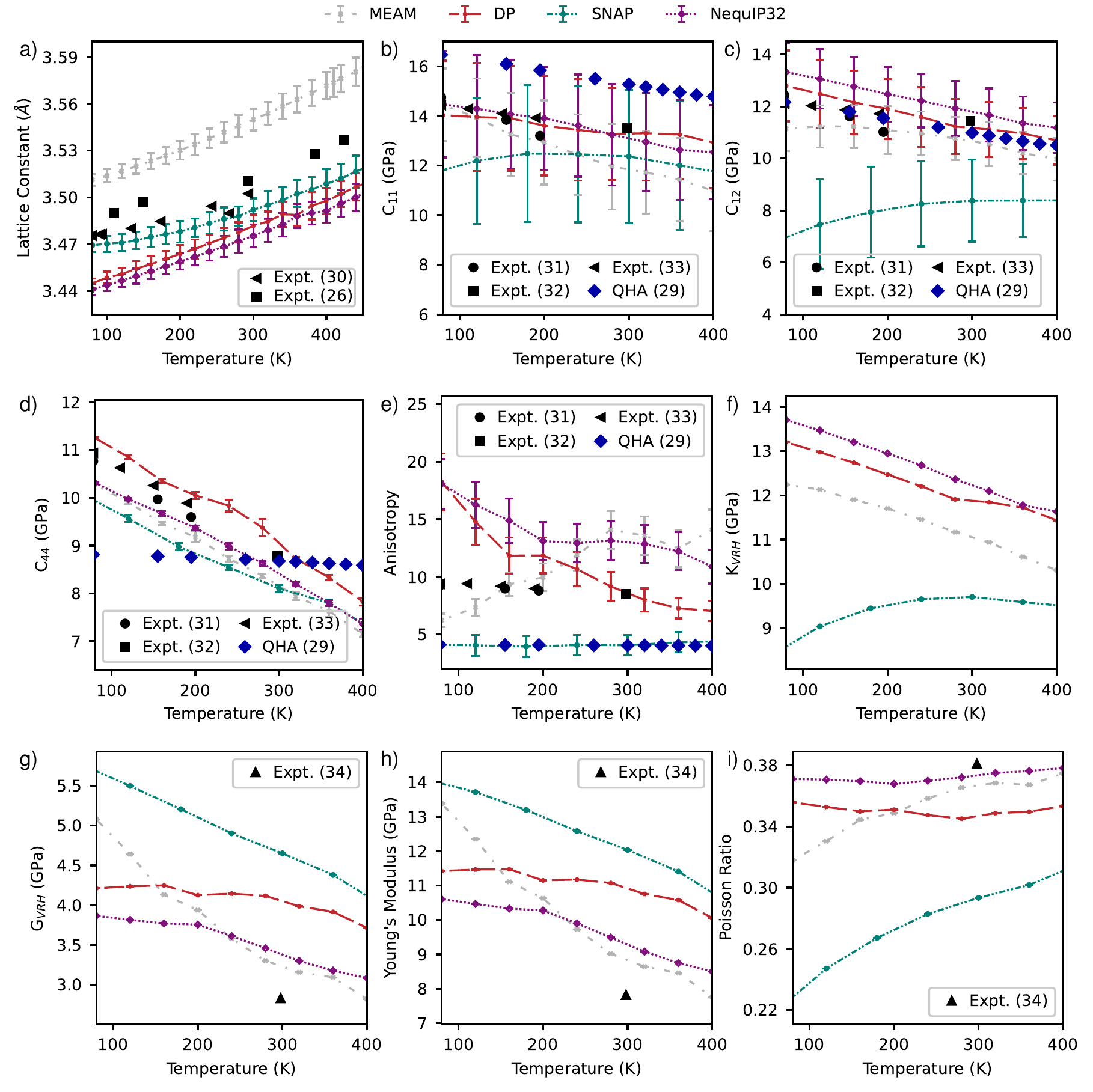}
\caption{Various bulk mechanical properties of lithium as a function of temperature. The NequIP and DP potentials in this work perform well across all bulk properties for which there is significant experimental data while MEAM performs reasonably well but slightly worse and SNAP performs poorly. We also compare the predictions to the Quasiharmonic Approximation (QHA) by  \cite{xu_enhanced_2017} and demonstrate the limitation of the QHA approach in finite temperature elastic constant prediction, especially for C$_{44}$. Error bars in the lattice constant (a) are the standard deviation due to fluctuations in the volume of the NPT simulation. Error bars in the elastic constants (b-d) are standard errors from the fitting of stress-strain curves and errorbars in the anisotropy (e) are propagated from errors in the elastic constants.}
\label{fig:elastic}
\end{figure*}
\nocite{owen_x-ray_1954}
\nocite{c_nash_single-crystal_1959}
\nocite{trivisonno_elastic_1961}
\nocite{slotwinski_temperature_1969}
\nocite{masias_elastic_2019}
We estimate a number of mechanical properties using MD implemented in the Large-scale Atomic/Molecular Massively Parallel Simulator (LAMMPS) using the potentials considered in this work. The results are used to predict key parameters in the design of Lithium Metal Solid  State Batteries (LMSSBs) such as the bulk and shear modulus as a function of temperature.  

The temperature dependence of the lattice constant of BCC lithium is shown in \fref{fig:lattice_constant}. Starting from the conventional equilibrium BCC structure in a 6x6x6 supercell at 0K for each potential, an NPT simulation is used to raise the temperature at a heating rate of 0.01K/timestep and then run at equilibrium for 100,000 timesteps with 1fs/timestep at temperature $T$ and zero external stress allowing the volume of the simulation box to change while keeping the unit cell orthorhombic. The average box size for the last 80,000 timesteps sampled every 100 timesteps can be used to extract the average lattice constant as a function of temperature.

All the MLIPs and MEAM perform well with less than 1.5\% error for the lattice constant with a slight underestimation for the MLIPs. MEAM overestimates the potential as the 0K parameters are fit to match room temperature result of 3.50\AA{} \cite{kim_atomistic_2012}. The temperature range considered was chosen because there exists a Martensitic transition into a FCC structure at 78K and the melting point of lithium is 450K \cite{beg_temperature_1976}. NequIP, MEAM and DP predict a strongly linear dependence of the lattice constant with temperature whereas SNAP predicts a slightly increasing thermal expansion coefficient corresponding to an increasing slope of the lattice constant with temperature. There is not enough experimental data above room temperature to determine which behavior is most likely. Such experiments may be challenging due to plasticity as they would be performed at a high homologous temperature, greater than 0.7 times the melting point for lithium.

The elastic response of single crystal BCC lithium, particularly at microscopic scales is key to the design of LMSSBs. The bulk and shear modulus are key parameters in the model of Monroe and Newman as well as Ahmad and Viswanathan used to predict stability against the formation of dendrites \cite{monroe_impact_2005,
 ahmad_stability_2017}. Due to lithium's low melting point, it is a soft material whose mechanical response can have unique properties near room temperature. The interplay between elastic and plastic regimes has been a topic of study \cite{xu_enhanced_2017, masias_elastic_2019, wang_mechanical_2019}. Here, we calculate the elastic constants of single crystal BCC lithium as a function of temperature. Xu et al. measured the elastic constants of lithium nanopillars and proceeded to calculate bulk elastic constants using a Quasiharmonic approximation (QHA) within DFT \cite{xu_enhanced_2017}. They found that the QHA performed poorly hence the need for simulations at the fidelity of AIMD. 

We perform the calculations  for the elastic constants $C_{11}$, $C_{12}$ and $C_{44}$ by fitting stress-strain curves after applying two different types of strain, an orthorhombic and monoclinic strain and perform 1ns long NVT simulations following the prescription by Zhang et al. \cite{zhang_prediction_2020} to save computational cost.  All other elastic constants for cubic crystals can be derived from these three using well known formulae \cite{nye_physical_1985} implemented in pymatgen \cite{jain_commentary_2013}.

The predictions of the components of the $C_{11}$, $C_{12}$ and $C_{44}$, the Universal Anisotropy, Voigt-Reuss-Hill averaged bulk and shear moduli ($K_{\text{VRH}}$ and $G_{\text{VRH}}$ respectively) as well as the Young's modulus and Poisson Ratio are plotted as a function of temperature in \fref{fig:elastic}. The Voigt-Reuss-Hill (VRH) average is a reliable method for predicting polycrystalline elastic moduli given the relevant single crystal elastic constants while being very easy to compute \cite{nye_physical_1985}.

As shown in \fref{fig:elastic}, the MLIPs in this work reproduce the experimental results well, significantly outperforming the QHA, SNAP and MEAM in reproducing experimental results for single crystal lithium and in the prediction of VRH averaged quantities.

The QHA is the only model that fails to predict the $C_{44}$ qualitatively accurately underestimating the dependence of $C_{44}$ as a function of temperature, potentially due to the assumed independence of vibrational modes in each spatial dimension. SNAP performs very poorly on $C_{11}$, $C_{12}$, predicting that they increase before decreasing with temperature, inconsistent with all experimental and previous computational results. The properties derived from these constants accumulate errors therefore SNAP is not suitable for calculations that rely on stress prediction. MEAM performs generally well on all predictions as it was specifically fit to elastic constants but predicts an increasing anisotropy with temperature different from all the other potentials and experiment. 

DP and NequIP32 are in excellent agreement with experimental results, consistently within 10\% or less of the experimental results for $C_{11}$, $C_{12}$ and $C_{44}$  and with matching qualitative behavior. NequIP32 performs slightly better than DP, likely due to the more accurate stress predictions, but both potentials are clearly suitable for prediction of bulk elastic properties. The only exception is the Anisotropy which is overestimated by DP and NequIP32 at low temperatures and shows much larger decrease with increasing temperature than experimental predictions. 

Overall, the DP and NequIP32 potentials in this work are the most accurate potentials with which to calculate bulk and elastic phenomena for BCC lithium.

\subsection{Adsorption and Surface Diffusion Barriers}\label{sec:adsorption}

\begin{figure*}[!tbh]
\centering
\labelphantom{fig:pes}
\labelphantom{fig:bep}
\labelphantom{fig:es_barrier}
\includegraphics[width=\textwidth]{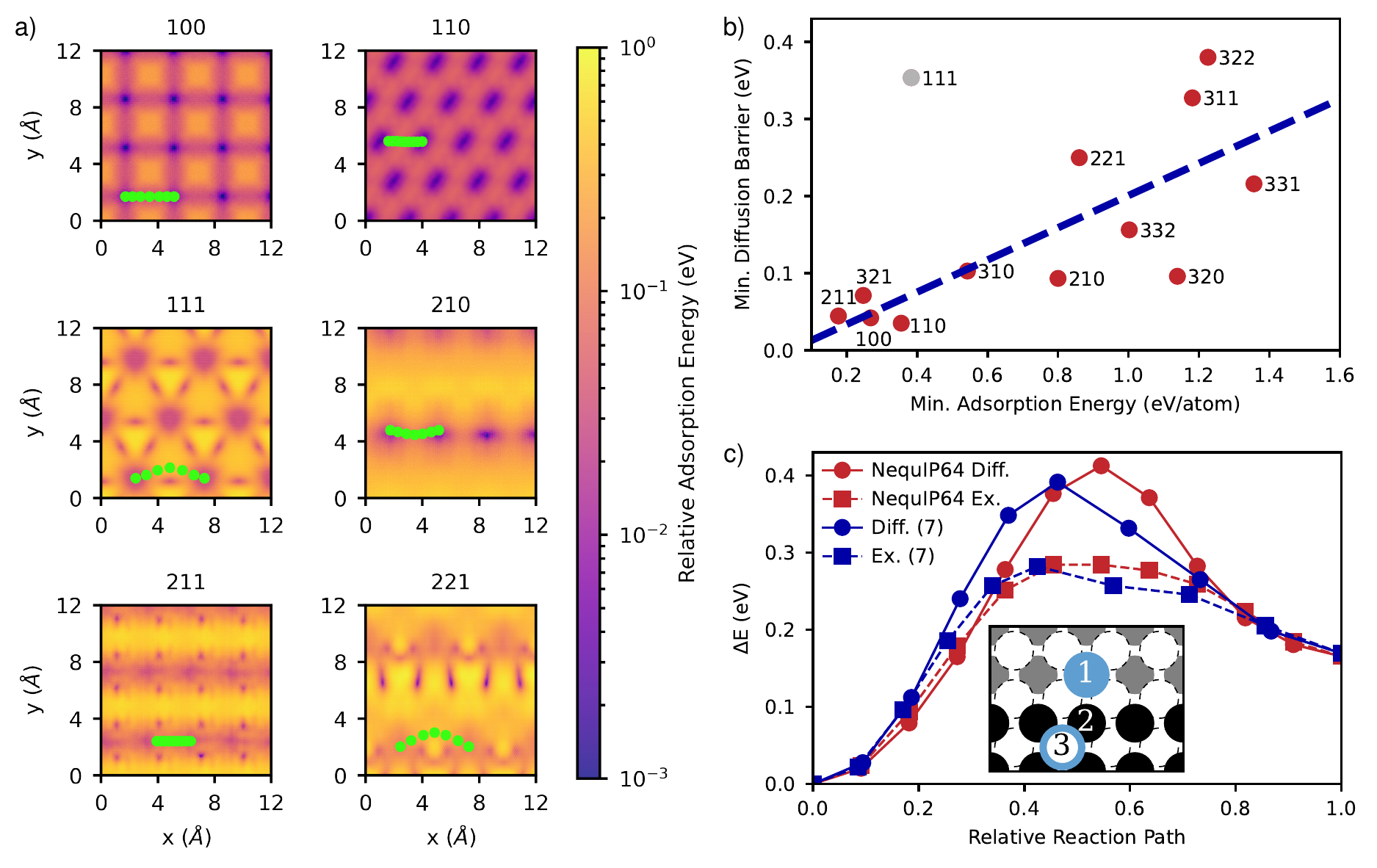}
\caption{Various surface properties calculated using the NequIP64 potential. a) Demonstration of a BEP correlation between the adsorption energy and the minimum diffusion barrier of each facet. The (111) surface was not included in the fitting of the dotted line. b) Calculation of an Erlich-Schwöebel barrier matching results in the literature. c) A variety of Surface Potential Energy Surfaces colored by the Relative Adsorption Energy show the the different geometries of adsorption sites and the minimum diffusion barrier paths adatoms can take when diffusing from one surface unit cell to the next.}
\label{fig:surfaces}
\end{figure*}

Another set of key properties for modeling lithium in the context of batteries involves surface properties such as adsorption and diffusion across the surface. There are a number of studies that perform these simulations using DFT but they are often forced to limit the surface area of the slabs, the miller indices considered and the number of layers in their slab models \cite{jackle_microscopic_2014, jackle_self-diffusion_2018, pande_computational_2019}. In our experiments, we found that it was impossible to converge the number of layers in the calculation of adsorption energy to within $\sim$1meV/atom, even for the lowest miller indices, with less than 10 layers. 
Sun et al. noted that the poor convergence of surface energies was likely due to inconsistencies in the Brillouin zone sampling of the clean slabs, bulk references and adsorbed slabs \cite{sun_efficient_2013}. They introduced a workaround where a slab-consistent bulk reference is used for faster convergence with fewer layers for surface energy. However, it is not clear how to create a bulk reference consistent with an adsorbed surface hence a need for the surface energy to converge even with a single atom bulk reference for adsorption calculations. Since the MLIP is not limited as much as DFT in the number of layers, we can adequately converge the number of layers, typically on the order of tens of layers even with large surface areas.

For DFT calculations of adsorption energies and surface diffusion barriers, the workaround is to fix a few layers, usually 2 at the bottom of a slab with 4-6 layers in total, artificially imposing a bulk environment close to the surface \cite{sholl_density_2009}. This limits the possible facets which can be considered to the very lowest few miller indices, typically up to 211. This is hampered as well by the failure of convergence with so few layers. With the MLIPs, the size of the slab model is less restricted thus we construct slabs of at least 14\AA{} in thickness with the first 6\AA{} of layers fixed consistent with the exposed facet onto which an adsorbate is placed in all our calculations using pymatgen.

For all results in this section, we use the most accurate NequIP64 potential. The DP potential also performs well on all the properties but to a lesser extent.

We plot the Surface Potential Energy Surface (SPES) for various miller indices in \fref{fig:pes}. The SPES is calculated by adsorbing a lithium atom on the relevant facet at various positions in the surface unit cell of a 4x4 surface supercell and allowing the adatom to relax only in the perpendicular direction (z direction) to the surface while allowing all other atoms (except the 6 fixed layers at the bottom of the slab) to relax in any direction. 
The adsorption energy ($E_{\text{ads}}$) is calculated using an unconventional method that uses a bulk reference for the adatom as 
\begin{align*}
    E_{\text{ads}} = E_{\text{adsorbed slab}} - E_{\text{clean slab}} - E_{\text{bulk atom}}.
\end{align*}
The bulk reference is chosen as it is more reliably predicted by all the potentials since no single atom data was added to the training datasets and we are only interested in relative adsorption energies and activation barriers. Relative Adsorption Energy is used in the SPES i.e. the energy above the lowest adsorption energy for that particular facet. The SPES shows the diverse geometries and distributions of adsorption sites that can arise with exposure of different facets and can potentially be used for thermodynamic lattice models of adsorption and surface diffusion properties. SPESs for higher miller indices can be found in the Supporting Information \siref{}.

Using the same slab models as those used for adsorption energies, we also calculate the surface diffusion barriers as an adatom diffuses from one surface unit cell to the next using the Nudged Elastic Band (NEB) method with a spring constant of 0.1eV/\AA{} and 7 images as implemented in ASE \cite{henkelman_climbing_2000}. The paths with the lowest diffusion barriers identified by the NEB method are shown on the SPES in \fref{fig:pes}. We find that for higher miller indices, the lithium atoms are more likely to diffuse via one-dimensional channels formed by step edges as there are higher coordination numbers in this environment and hence lower adsorption energies in agreement with Gaissmaier et al. Only (100) and (110) and (111) have equally high diffusion barriers in two dimensions \cite{gaissmaier_first_2019}.

In \fref{fig:bep}, we plot the adsorption energy versus the lowest surface diffusion barrier for that facet and show that there exists a linear correlation between the adsorption energy and the surface diffusion barrier. This is an example of a Bell-Evans-Polanyi (BEP) principle, a general framework that notes that the difference in activation energy between two reactions of the same family is roughly proportional to the difference of their enthalpy of reaction \cite{bell_theory_1997, evans_further_1936}. In the context of adsorption energies and surface diffusion barriers, Pande et al. found a BEP relation relating the adsorption energies on low miller index facets for different metals and alloys and the surface diffusion barriers on those surfaces \cite{pande_computational_2019}. They argued that the BEP relation can be used as a powerful screening tool that avoids expensive NEB calculations. We show that there also exists a BEP relation for different miller indices of the same material which can similarly be used for screening purposes in cases where the NEB method is too costly. We note that the  (111) surface is an outlier to the BEP relation, likely due to it's very high surface energy as it is not a close-packed surface for the BCC crystal structure and all the surface atoms are under-coordinated.

We also predict an example of an Erlich-Schwöebel barrier, a descriptor used in determining the likelihood of an adatom to diffuse up a step-edge \cite{schwoebel_step_1966}. The activation energy for this process is increased as the adatom goes from a region of high coordination along the step-edge to lower coordination on top of the step. The Erlich-Schwöebel activation barrier (E$_S$) is defined as $$ E_S = E_{ESB} - E_T $$ where E$_{\text{ESB}}$ is the increased barrier due to having to diffuse up a step-edge and E$_T$ is the activation energy for diffusion without any steps \cite{gaissmaier_first_2019}. We consider two mechanisms for step-up diffusion on the most stable (100) surface and show in \fref{fig:es_barrier} that NequIP64 reproduces the results by Gaissmaier et al. to within 0.01eV for the activation energy for both mechanisms\cite{gaissmaier_first_2019}. A schematic to explain the two mechanisms is shown in \fref{fig:es_barrier}. The first mechanism is when the adatom directly diffuses (Diff.) over the step-edge (1$\rightarrow$3) and the second is by exchanging (Ex.) with an atom in the step (1$\rightarrow$2, 2$\rightarrow$3).

Overall the NequIP64 potential has shown very good accuracy in reproducing and predicting surface properties and has allowed us to predict properties that are computationally infeasible using DFT. 

\section*{Discussion: Implications for Li-metal Battery Design}
The calculations enabled by the lithium MLIPs derived in this work now allow refining criterion for morphological stability associated with dendrite formation described above.  Monroe and Newman, subsequently, Ahmad and Viswanathan showed that room temperature shear modulus, anisotropy of lithium are critical factors in determining the stability criterion for dendrite suppression and thereby the required shear modulus of the solid electrolyte \cite{monroe_impact_2005, ahmad_stability_2017}.  In this work, we show that the temperature-dependent response of the mechanical properties is much stronger than that previously predicted using the QHA.  The room-temperature modulus, as predicted by NequIP32 is around 3.5 GPa, a reduction of about $\sim$30\% from that at 0K using the QHA, drastically modifying the stability criterion, thereby modifying the required shear modulus of the solid electrolyte needed to suppress dendrite formation by a similar $\sim$30\%.

Jäckle et al. and Gaissmaier et al. only probed surface diffusion barriers for low Miller index facets, tractable using DFT calculations \cite{jackle_self-diffusion_2018, gaissmaier_first_2019}. Morphological protrusions can often locally have high Miller index domains with a high degree of under-coordination such as the curved surface of a roughly cylindrical dendrite. Using the lithium MLIP developed here, we show the existence of a BEP relation, indicating that high Miller indices which typically have higher lithium adsorption energy due to under-coordination have much larger surface diffusion barriers.  For the (110) and (211) facets which appear in the Wulff construction, the diffusion rates $\nu\propto\exp{(-\Delta E/k_BT)}$ at 300K are 1-2 times slower than on the (100) facet with the lowest barrier. The exception is the (111) surface which has an anomalously large surface diffusion barrier with respect to the BEP relation. Other facets which do not appear in the Wulff construction all have higher diffusion barriers with $\nu>$4 times that of the (100) surface or many orders of magnitude more, even along preferential one-dimensional channels observed in the SPES. 

Another implication is that glassy phases which have been shown to improve bulk conductivity \cite{wang_glassy_2020}, are likely to possess lower surface diffusion limiting the charging rate when glassy phases of lithium are formed.  Surface diffusion in glassy phases is further hampered by the lack of long range order as lithium is deposited which would randomly orient the preferential one-dimensional channels with high adsorption energy thereby trapping atoms in deep potential wells. We therefore expect that the combined effect is the proliferation of local instabilities under fast charging conditions before the surface equilibrates to (100) and (110) facets with faster diffusion in two dimensions which would reduce dendrite growth. 

We believe that the lithium MLIP and training dataset developed here will enable the calculation of void formation, creep behavior and various other meso-scale properties previously not tractable using atomistic simulations with significantly improved accuracy thereby allowing refined material properties under different conditions.

\section{Methods}
In this section, we describe the model architectures, data generation and simulation details.

\subsection*{Machine Learning Interaction Potentials}
MLIPs are parametric models that can be trained on a set of atomic configurations given in terms of coordinates $\{\mathbf{R}\}$ and labelled with the corresponding energies, forces and stresses for that configuration from a high fidelity source such as DFT. 

Typically, the total energy of an atomic configuration is modeled as a sum of atomic energies as in \eqref{eq:energy} i.e.

\begin{equation*}\label{eq:energy}
    E(\{\mathbf{R}\}) = \sum\limits_i\varepsilon_i.
\end{equation*} 
The $i$-th atom contributes an energy ($\varepsilon_i$) that is learned from the geometry centered on atom $i$. The forces on each atom and the virial stresses can then be calculated as the derivatives of the total energy with respect to the atomic positions and the cell dimensions respectively using automatic differentiation. It is essential, that the predicted energies, forces and stress are appropriately equivariant with respect to translations, rotations and permutations of atoms of the same species. The key difference between different potentials is how they implement this equivariance as described for Deep Potential and NequIP.
\\
\textit{Deep Potential} - Deep Potential provides a trainable, fully local framework for training MLIPs that can predict energy, forces and stresses of a particular configuration of atoms. In this work, we choose the se\textunderscore e3 descriptor constructed from radial and angular information of atomic environments. This descriptor is invariant to the symmetries and not trainable. The neural network architecture is that of Residual Neural Network \cite{zhang_deep_2018}. 
\\
\textit{NequIP} - The Neural Equivariant Interatomic Potential (NequIP) is an E(3)-equivariant message passing interatomic potential that was shown to demonstrate state-of-the-art accuracy, sample efficiency, and transferability on a variety of materials systems at the time of writing \cite{batzner_e3-equivariant_2022, musaelian_learning_2022, batatia_design_2022}. While conventional interatomic potentials operate on invariant descriptors of the materials systems, such as distances and angles, NequIP directly operates on relative interatomic positions $\vec{r}_{ij}$ represented as a graph and leverages latent features comprised of not only scalar, but also vector and higher-order tensor features. 

\subsection*{DFT Data}

\begin{figure*}[tbhp]
\centering
\includegraphics[width=\linewidth]{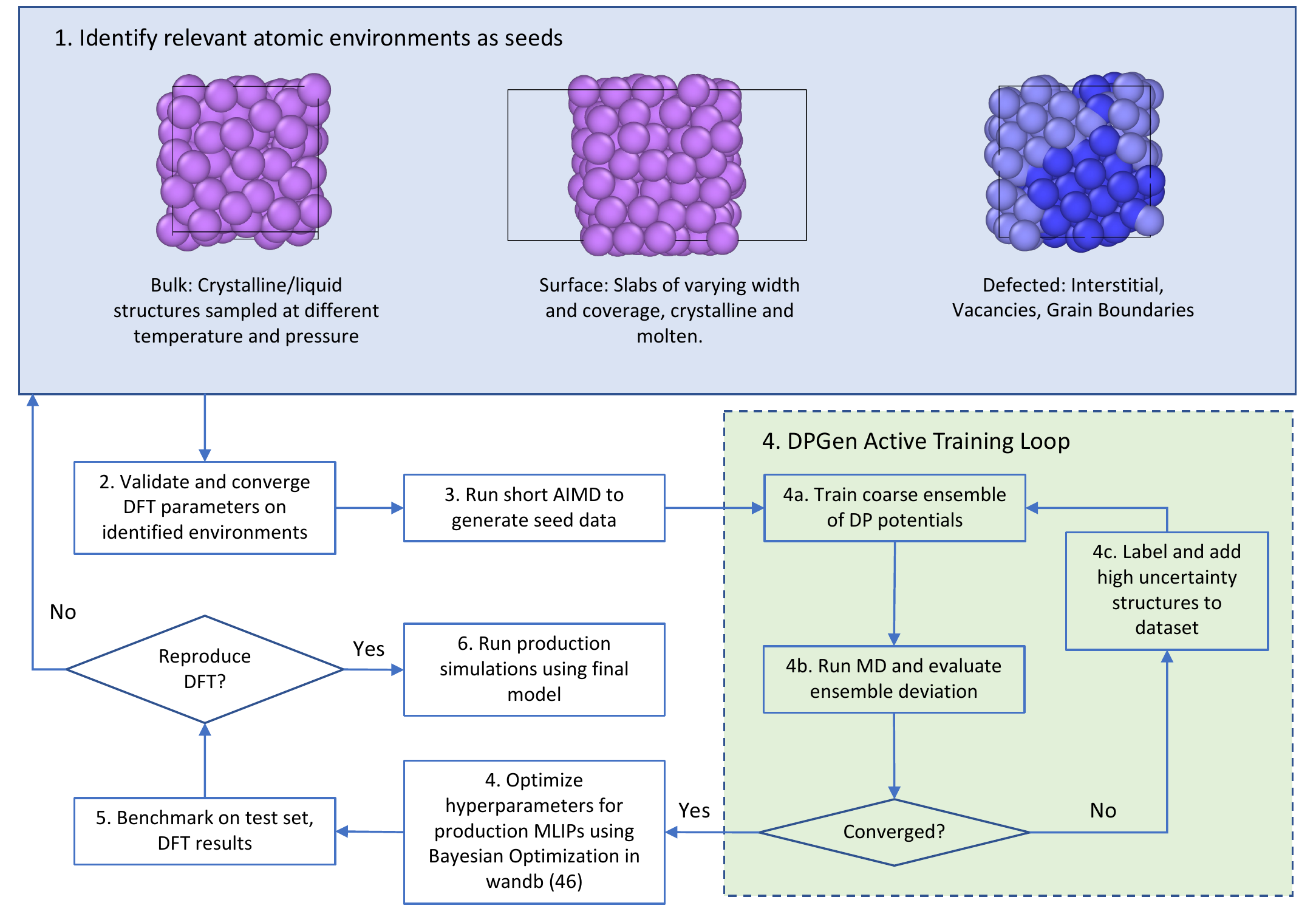}
\caption{Schematic showing the process by which data was generated and potentials were validated.}
\label{fig:workflow}
\end{figure*}
\nocite{wandb}

A schematic of the procedure used to generate data is shown in \fref{fig:workflow}. All data used to train the MLIPs was generated using Density Functional Theory with the same parameters across the entire dataset. The parameters chosen were such that the brillouin zone sampling density, plane wave and density cutoffs were converged to <1meV/atom to ensure consistency and give a concrete estimate of statistical noise in the data. DFT calculations were done using Quantum Espresso \cite{giannozzi_quantum_2009} within the Generalized Gradient Approximation using the Perdew-Burke-Eizenhoff exchange correlation functional \cite{perdew_generalized_1997} and the Projector Augemented Wave approach \cite{blochl_projector_1994} with a plane wave cutoff energy of 1360eV. We used the pseudopotential Li.pbe-s-kjpaw\_psl.1.0.0.UPF from
http://www.quantum-espresso.org. A uniform Brillouin Zone spacing of 0.02$\AA^{-1}$ with a Monkhorst-Pack \cite{monkhorst_special_1976} sampling procedure was used. To help with convergence of the the Fermi surface, Methfessel-Paxton \cite{methfessel_high-precision_1989} smearing using a smearing width of 0.27eV was chosen.

The data used to train the potentials was generated using an active learning approach as implemented in the DPGen package \cite{zhang_dp-gen_2020}. A variety of seed structures including bulk crystals, monovacant, mono-interstitial, clean surfaces and surfaces with varying coverage up to a maximum of 128 atoms. The seed structures were  used as starting structures for isobaric, isothermal (NPT) MD simulations. The MD simulations were performed sequentially on a grid with temperatures of 50K, 300K, 450K and 900K i.e. double the melting temperature and pressures of  0GPa, 0.1GPa, 1GPa and 10GPa. With each active learning step in the sequence, data from previous steps was incorporated into subsequent steps to improve accuracy and flag new structures with high uncertainty for labeling. The final dataset used to train production potentials was split into 4548 structures in the training set and 505 configurations in the test set. 
Convergence of the dataset and potentials was determined using RMSEs and accurate property prediction across phenomena in all the different classes of configurations considered relevant for a lithium potential of the entire phase space.

\clearpage
\bibliography{refs}

\providecommand{\latin}[1]{#1}
\makeatletter
\providecommand{\doi}
  {\begingroup\let\do\@makeother\dospecials
  \catcode`\{=1 \catcode`\}=2 \doi@aux}
\providecommand{\doi@aux}[1]{\endgroup\texttt{#1}}
\makeatother
\providecommand*\mcitethebibliography{\thebibliography}
\csname @ifundefined\endcsname{endmcitethebibliography}
  {\let\endmcitethebibliography\endthebibliography}{}
\begin{mcitethebibliography}{53}
\providecommand*\natexlab[1]{#1}
\providecommand*\mciteSetBstSublistMode[1]{}
\providecommand*\mciteSetBstMaxWidthForm[2]{}
\providecommand*\mciteBstWouldAddEndPuncttrue
  {\def\EndOfBibitem{\unskip.}}
\providecommand*\mciteBstWouldAddEndPunctfalse
  {\let\EndOfBibitem\relax}
\providecommand*\mciteSetBstMidEndSepPunct[3]{}
\providecommand*\mciteSetBstSublistLabelBeginEnd[3]{}
\providecommand*\EndOfBibitem{}
\mciteSetBstSublistMode{f}
\mciteSetBstMaxWidthForm{subitem}{(\alph{mcitesubitemcount})}
\mciteSetBstSublistLabelBeginEnd
  {\mcitemaxwidthsubitemform\space}
  {\relax}
  {\relax}

\bibitem[Wang \latin{et~al.}(2020)Wang, Cui, Chu, and Wu]{wang_lithium_2020}
Wang,~R.; Cui,~W.; Chu,~F.; Wu,~F. Lithium metal anodes: {Present} and future.
  \emph{Journal of Energy Chemistry} \textbf{2020}, \emph{48}, 145--159, DOI:
  \doi{10.1016/j.jechem.2019.12.024}\relax
\mciteBstWouldAddEndPuncttrue
\mciteSetBstMidEndSepPunct{\mcitedefaultmidpunct}
{\mcitedefaultendpunct}{\mcitedefaultseppunct}\relax
\EndOfBibitem
\bibitem[Zhu \latin{et~al.}(2020)Zhu, Pande, Li, Wen, Pan, Wang, Ma,
  Viswanathan, and Chiang]{zhu_design_2020}
Zhu,~Y.; Pande,~V.; Li,~L.; Wen,~B.; Pan,~M.~S.; Wang,~D.; Ma,~Z.-F.;
  Viswanathan,~V.; Chiang,~Y.-M. Design principles for self-forming interfaces
  enabling stable lithium-metal anodes. \emph{Proceedings of the National
  Academy of Sciences} \textbf{2020}, \emph{117}, 27195--27203, DOI:
  \doi{10.1073/pnas.2001923117}, Publisher: Proceedings of the National Academy
  of Sciences\relax
\mciteBstWouldAddEndPuncttrue
\mciteSetBstMidEndSepPunct{\mcitedefaultmidpunct}
{\mcitedefaultendpunct}{\mcitedefaultseppunct}\relax
\EndOfBibitem
\bibitem[Lin \latin{et~al.}(2017)Lin, Liu, and Cui]{lin_reviving_2017}
Lin,~D.; Liu,~Y.; Cui,~Y. Reviving the lithium metal anode for high-energy
  batteries. \emph{Nature Nanotechnology} \textbf{2017}, \emph{12}, 194--206,
  DOI: \doi{10.1038/nnano.2017.16}, Number: 3 Publisher: Nature Publishing
  Group\relax
\mciteBstWouldAddEndPuncttrue
\mciteSetBstMidEndSepPunct{\mcitedefaultmidpunct}
{\mcitedefaultendpunct}{\mcitedefaultseppunct}\relax
\EndOfBibitem
\bibitem[Monroe and Newman(2005)Monroe, and Newman]{monroe_impact_2005}
Monroe,~C.; Newman,~J. The {Impact} of {Elastic} {Deformation} on {Deposition}
  {Kinetics} at {Lithium}/{Polymer} {Interfaces}. \emph{Journal of The
  Electrochemical Society} \textbf{2005}, \emph{152}, A396, DOI:
  \doi{10.1149/1.1850854}, Publisher: IOP Publishing\relax
\mciteBstWouldAddEndPuncttrue
\mciteSetBstMidEndSepPunct{\mcitedefaultmidpunct}
{\mcitedefaultendpunct}{\mcitedefaultseppunct}\relax
\EndOfBibitem
\bibitem[Ahmad and Viswanathan(2017)Ahmad, and
  Viswanathan]{ahmad_stability_2017}
Ahmad,~Z.; Viswanathan,~V. Stability of {Electrodeposition} at {Solid}-{Solid}
  {Interfaces} and {Implications} for {Metal} {Anodes}. \emph{Physical Review
  Letters} \textbf{2017}, \emph{119}, 056003, DOI:
  \doi{10.1103/PhysRevLett.119.056003}, Publisher: American Physical
  Society\relax
\mciteBstWouldAddEndPuncttrue
\mciteSetBstMidEndSepPunct{\mcitedefaultmidpunct}
{\mcitedefaultendpunct}{\mcitedefaultseppunct}\relax
\EndOfBibitem
\bibitem[Jäckle \latin{et~al.}(2018)Jäckle, Helmbrecht, Smits, Stottmeister,
  and Groß]{jackle_self-diffusion_2018}
Jäckle,~M.; Helmbrecht,~K.; Smits,~M.; Stottmeister,~D.; Groß,~A.
  Self-diffusion barriers: possible descriptors for dendrite growth in
  batteries? \emph{Energy \& Environmental Science} \textbf{2018}, \emph{11},
  3400--3407, DOI: \doi{10.1039/C8EE01448E}, Publisher: The Royal Society of
  Chemistry\relax
\mciteBstWouldAddEndPuncttrue
\mciteSetBstMidEndSepPunct{\mcitedefaultmidpunct}
{\mcitedefaultendpunct}{\mcitedefaultseppunct}\relax
\EndOfBibitem
\bibitem[Gaissmaier \latin{et~al.}(2019)Gaissmaier, Fantauzzi, and
  Jacob]{gaissmaier_first_2019}
Gaissmaier,~D.; Fantauzzi,~D.; Jacob,~T. First principles studies of
  self-diffusion processes on metallic lithium surfaces. \emph{The Journal of
  Chemical Physics} \textbf{2019}, \emph{150}, 041723, DOI:
  \doi{10.1063/1.5056226}, Publisher: American Institute of Physics\relax
\mciteBstWouldAddEndPuncttrue
\mciteSetBstMidEndSepPunct{\mcitedefaultmidpunct}
{\mcitedefaultendpunct}{\mcitedefaultseppunct}\relax
\EndOfBibitem
\bibitem[Li \latin{et~al.}(2017)Li, Li, Pei, Yan, Sun, Wu, Joubert, Chin, Koh,
  Yu, Perrino, Butz, Chu, and Cui]{li_atomic_2017}
Li,~Y.; Li,~Y.; Pei,~A.; Yan,~K.; Sun,~Y.; Wu,~C.-L.; Joubert,~L.-M.; Chin,~R.;
  Koh,~A.~L.; Yu,~Y.; Perrino,~J.; Butz,~B.; Chu,~S.; Cui,~Y. Atomic structure
  of sensitive battery materials and interfaces revealed by cryo–electron
  microscopy. \emph{Science} \textbf{2017}, \emph{358}, 506--510, DOI:
  \doi{10.1126/science.aam6014}, Publisher: American Association for the
  Advancement of Science\relax
\mciteBstWouldAddEndPuncttrue
\mciteSetBstMidEndSepPunct{\mcitedefaultmidpunct}
{\mcitedefaultendpunct}{\mcitedefaultseppunct}\relax
\EndOfBibitem
\bibitem[Kresse and Furthmüller(1996)Kresse, and
  Furthmüller]{kresse_efficiency_1996}
Kresse,~G.; Furthmüller,~J. Efficiency of ab-initio total energy calculations
  for metals and semiconductors using a plane-wave basis set.
  \emph{Computational Materials Science} \textbf{1996}, \emph{6}, 15--50, DOI:
  \doi{10.1016/0927-0256(96)00008-0}\relax
\mciteBstWouldAddEndPuncttrue
\mciteSetBstMidEndSepPunct{\mcitedefaultmidpunct}
{\mcitedefaultendpunct}{\mcitedefaultseppunct}\relax
\EndOfBibitem
\bibitem[Behler and Csányi(2021)Behler, and Csányi]{behler_machine_2021}
Behler,~J.; Csányi,~G. Machine learning potentials for extended systems: a
  perspective. \emph{The European Physical Journal B} \textbf{2021}, \emph{94},
  142, DOI: \doi{10.1140/epjb/s10051-021-00156-1}\relax
\mciteBstWouldAddEndPuncttrue
\mciteSetBstMidEndSepPunct{\mcitedefaultmidpunct}
{\mcitedefaultendpunct}{\mcitedefaultseppunct}\relax
\EndOfBibitem
\bibitem[Batzner \latin{et~al.}(2022)Batzner, Musaelian, Sun, Geiger, Mailoa,
  Kornbluth, Molinari, Smidt, and Kozinsky]{batzner_e3-equivariant_2022}
Batzner,~S.; Musaelian,~A.; Sun,~L.; Geiger,~M.; Mailoa,~J.~P.; Kornbluth,~M.;
  Molinari,~N.; Smidt,~T.~E.; Kozinsky,~B. E(3)-equivariant graph neural
  networks for data-efficient and accurate interatomic potentials. \emph{Nature
  Communications} \textbf{2022}, \emph{13}, 2453, DOI:
  \doi{10.1038/s41467-022-29939-5}, Number: 1 Publisher: Nature Publishing
  Group\relax
\mciteBstWouldAddEndPuncttrue
\mciteSetBstMidEndSepPunct{\mcitedefaultmidpunct}
{\mcitedefaultendpunct}{\mcitedefaultseppunct}\relax
\EndOfBibitem
\bibitem[Schütt \latin{et~al.}(2018)Schütt, Sauceda, Kindermans, Tkatchenko,
  and Müller]{schutt_schnet_2018}
Schütt,~K.~T.; Sauceda,~H.~E.; Kindermans,~P.-J.; Tkatchenko,~A.;
  Müller,~K.-R. {SchNet} – {A} deep learning architecture for molecules and
  materials. \emph{The Journal of Chemical Physics} \textbf{2018}, \emph{148},
  241722, DOI: \doi{10.1063/1.5019779}, Publisher: American Institute of
  Physics\relax
\mciteBstWouldAddEndPuncttrue
\mciteSetBstMidEndSepPunct{\mcitedefaultmidpunct}
{\mcitedefaultendpunct}{\mcitedefaultseppunct}\relax
\EndOfBibitem
\bibitem[Cheng \latin{et~al.}(2020)Cheng, Mazzola, Pickard, and
  Ceriotti]{cheng_evidence_2020}
Cheng,~B.; Mazzola,~G.; Pickard,~C.~J.; Ceriotti,~M. Evidence for supercritical
  behaviour of high-pressure liquid hydrogen. \emph{Nature} \textbf{2020},
  \emph{585}, 217--220, DOI: \doi{10.1038/s41586-020-2677-y}, Number: 7824
  Publisher: Nature Publishing Group\relax
\mciteBstWouldAddEndPuncttrue
\mciteSetBstMidEndSepPunct{\mcitedefaultmidpunct}
{\mcitedefaultendpunct}{\mcitedefaultseppunct}\relax
\EndOfBibitem
\bibitem[Freitas and Cao(2022)Freitas, and Cao]{freitas_machine-learning_2022}
Freitas,~R.; Cao,~Y. Machine-learning potentials for crystal defects. \emph{MRS
  Communications} \textbf{2022}, \emph{12}, 510--520, DOI:
  \doi{10.1557/s43579-022-00221-5}\relax
\mciteBstWouldAddEndPuncttrue
\mciteSetBstMidEndSepPunct{\mcitedefaultmidpunct}
{\mcitedefaultendpunct}{\mcitedefaultseppunct}\relax
\EndOfBibitem
\bibitem[Owen \latin{et~al.}(2023)Owen, Torrisi, Xie, Batzner, Coulter,
  Musaelian, Sun, and Kozinsky]{owen_complexity_2023}
Owen,~C.~J.; Torrisi,~S.~B.; Xie,~Y.; Batzner,~S.; Coulter,~J.; Musaelian,~A.;
  Sun,~L.; Kozinsky,~B. Complexity of {Many}-{Body} {Interactions} in
  {Transition} {Metals} via {Machine}-{Learned} {Force} {Fields} from the
  {TM23} {Data} {Set}. 2023; \url{http://arxiv.org/abs/2302.12993},
  arXiv:2302.12993 [cond-mat, physics:physics]\relax
\mciteBstWouldAddEndPuncttrue
\mciteSetBstMidEndSepPunct{\mcitedefaultmidpunct}
{\mcitedefaultendpunct}{\mcitedefaultseppunct}\relax
\EndOfBibitem
\bibitem[Zuo \latin{et~al.}(2020)Zuo, Chen, Li, Deng, Chen, Behler, Csányi,
  Shapeev, Thompson, Wood, and Ong]{zuo_performance_2020}
Zuo,~Y.; Chen,~C.; Li,~X.; Deng,~Z.; Chen,~Y.; Behler,~J.; Csányi,~G.;
  Shapeev,~A.~V.; Thompson,~A.~P.; Wood,~M.~A.; Ong,~S.~P. Performance and
  {Cost} {Assessment} of {Machine} {Learning} {Interatomic} {Potentials}.
  \emph{The Journal of Physical Chemistry A} \textbf{2020}, \emph{124},
  731--745, DOI: \doi{10.1021/acs.jpca.9b08723}, Publisher: American Chemical
  Society\relax
\mciteBstWouldAddEndPuncttrue
\mciteSetBstMidEndSepPunct{\mcitedefaultmidpunct}
{\mcitedefaultendpunct}{\mcitedefaultseppunct}\relax
\EndOfBibitem
\bibitem[Zhang \latin{et~al.}(2018)Zhang, Han, Wang, Car, and
  E]{zhang_deep_2018}
Zhang,~L.; Han,~J.; Wang,~H.; Car,~R.; E,~W. Deep {Potential} {Molecular}
  {Dynamics}: {A} {Scalable} {Model} with the {Accuracy} of {Quantum}
  {Mechanics}. \emph{Physical Review Letters} \textbf{2018}, \emph{120},
  143001, DOI: \doi{10.1103/PhysRevLett.120.143001}\relax
\mciteBstWouldAddEndPuncttrue
\mciteSetBstMidEndSepPunct{\mcitedefaultmidpunct}
{\mcitedefaultendpunct}{\mcitedefaultseppunct}\relax
\EndOfBibitem
\bibitem[Jiao \latin{et~al.}(2022)Jiao, Lai, Zhao, Lu, Li, Xu, Jiang, He,
  Ouyang, Pan, Li, and Zheng]{jiao_self-healing_2022}
Jiao,~J.; Lai,~G.; Zhao,~L.; Lu,~J.; Li,~Q.; Xu,~X.; Jiang,~Y.; He,~Y.-B.;
  Ouyang,~C.; Pan,~F.; Li,~H.; Zheng,~J. Self-{Healing} {Mechanism} of
  {Lithium} in {Lithium} {Metal}. \emph{Advanced Science} \textbf{2022},
  \emph{9}, 2105574, DOI: \doi{10.1002/advs.202105574}, \_eprint:
  https://onlinelibrary.wiley.com/doi/pdf/10.1002/advs.202105574\relax
\mciteBstWouldAddEndPuncttrue
\mciteSetBstMidEndSepPunct{\mcitedefaultmidpunct}
{\mcitedefaultendpunct}{\mcitedefaultseppunct}\relax
\EndOfBibitem
\bibitem[Kim \latin{et~al.}(2012)Kim, Jung, and Lee]{kim_atomistic_2012}
Kim,~Y.-M.; Jung,~I.-H.; Lee,~B.-J. Atomistic modeling of pure {Li} and
  {Mg}–{Li} system. \emph{Modelling and Simulation in Materials Science and
  Engineering} \textbf{2012}, \emph{20}, 035005, DOI:
  \doi{10.1088/0965-0393/20/3/035005}\relax
\mciteBstWouldAddEndPuncttrue
\mciteSetBstMidEndSepPunct{\mcitedefaultmidpunct}
{\mcitedefaultendpunct}{\mcitedefaultseppunct}\relax
\EndOfBibitem
\bibitem[Luo \latin{et~al.}(2022)Luo, Zhang, Liu, Wang, Fan, Wang, Ma, Zhu, and
  Zhang]{luo_thermal_2022}
Luo,~S.; Zhang,~Y.; Liu,~X.; Wang,~Z.; Fan,~A.; Wang,~H.; Ma,~W.; Zhu,~L.;
  Zhang,~X. Thermal behavior of {Li} electrode in all-solid-state batteries and
  improved performance by temperature modulation. \emph{International Journal
  of Heat and Mass Transfer} \textbf{2022}, \emph{199}, 123450, DOI:
  \doi{10.1016/j.ijheatmasstransfer.2022.123450}\relax
\mciteBstWouldAddEndPuncttrue
\mciteSetBstMidEndSepPunct{\mcitedefaultmidpunct}
{\mcitedefaultendpunct}{\mcitedefaultseppunct}\relax
\EndOfBibitem
\bibitem[Wang \latin{et~al.}(2020)Wang, Pawar, Li, Ren, Zhang, Lu, Banerjee,
  Liu, Dufek, Zhang, Xiao, Liu, Meng, and Liaw]{wang_glassy_2020}
Wang,~X.; Pawar,~G.; Li,~Y.; Ren,~X.; Zhang,~M.; Lu,~B.; Banerjee,~A.; Liu,~P.;
  Dufek,~E.~J.; Zhang,~J.-G.; Xiao,~J.; Liu,~J.; Meng,~Y.~S.; Liaw,~B. Glassy
  {Li} metal anode for high-performance rechargeable {Li} batteries.
  \emph{Nature Materials} \textbf{2020}, 1--7, DOI:
  \doi{10.1038/s41563-020-0729-1}, Publisher: Nature Publishing Group\relax
\mciteBstWouldAddEndPuncttrue
\mciteSetBstMidEndSepPunct{\mcitedefaultmidpunct}
{\mcitedefaultendpunct}{\mcitedefaultseppunct}\relax
\EndOfBibitem
\bibitem[Batatia \latin{et~al.}(2022)Batatia, Batzner, Kovács, Musaelian,
  Simm, Drautz, Ortner, Kozinsky, and Csányi]{batatia_design_2022}
Batatia,~I.; Batzner,~S.; Kovács,~D.~P.; Musaelian,~A.; Simm,~G. N.~C.;
  Drautz,~R.; Ortner,~C.; Kozinsky,~B.; Csányi,~G. The {Design} {Space} of
  {E}(3)-{Equivariant} {Atom}-{Centered} {Interatomic} {Potentials}. 2022;
  \url{http://arxiv.org/abs/2205.06643}, arXiv:2205.06643 [cond-mat,
  physics:physics, stat]\relax
\mciteBstWouldAddEndPuncttrue
\mciteSetBstMidEndSepPunct{\mcitedefaultmidpunct}
{\mcitedefaultendpunct}{\mcitedefaultseppunct}\relax
\EndOfBibitem
\bibitem[Tran \latin{et~al.}(2016)Tran, Xu, Radhakrishnan, Winston, Sun,
  Persson, and Ong]{tran_surface_2016}
Tran,~R.; Xu,~Z.; Radhakrishnan,~B.; Winston,~D.; Sun,~W.; Persson,~K.~A.;
  Ong,~S.~P. Surface energies of elemental crystals. \emph{Scientific Data}
  \textbf{2016}, \emph{3}, 160080, DOI: \doi{10.1038/sdata.2016.80}, Number: 1
  Publisher: Nature Publishing Group\relax
\mciteBstWouldAddEndPuncttrue
\mciteSetBstMidEndSepPunct{\mcitedefaultmidpunct}
{\mcitedefaultendpunct}{\mcitedefaultseppunct}\relax
\EndOfBibitem
\bibitem[Sholl(2009)]{sholl_density_2009}
Sholl,~D.~S. \emph{Density functional theory a practical introduction}; Wiley:
  Hoboken, N.J, 2009\relax
\mciteBstWouldAddEndPuncttrue
\mciteSetBstMidEndSepPunct{\mcitedefaultmidpunct}
{\mcitedefaultendpunct}{\mcitedefaultseppunct}\relax
\EndOfBibitem
\bibitem[Musaelian \latin{et~al.}(2022)Musaelian, Batzner, Johansson, Sun,
  Owen, Kornbluth, and Kozinsky]{musaelian_learning_2022}
Musaelian,~A.; Batzner,~S.; Johansson,~A.; Sun,~L.; Owen,~C.~J.; Kornbluth,~M.;
  Kozinsky,~B. Learning {Local} {Equivariant} {Representations} for
  {Large}-{Scale} {Atomistic} {Dynamics}. \emph{arXiv:2204.05249 [cond-mat,
  physics:physics]} \textbf{2022}, arXiv: 2204.05249\relax
\mciteBstWouldAddEndPuncttrue
\mciteSetBstMidEndSepPunct{\mcitedefaultmidpunct}
{\mcitedefaultendpunct}{\mcitedefaultseppunct}\relax
\EndOfBibitem
\bibitem[Beg and Nielsen(1976)Beg, and Nielsen]{beg_temperature_1976}
Beg,~M.~M.; Nielsen,~M. Temperature {Dependence} of {Lattice} {Dynamics} of
  {Lithium} 7. \emph{Physical Review B (Condensed Matter and Materials
  Physics)} \textbf{1976}, \emph{14}, 4266--4273, DOI:
  \doi{10.1103/PhysRevB.14.4266}\relax
\mciteBstWouldAddEndPuncttrue
\mciteSetBstMidEndSepPunct{\mcitedefaultmidpunct}
{\mcitedefaultendpunct}{\mcitedefaultseppunct}\relax
\EndOfBibitem
\bibitem[Xie and Zhang(2011)Xie, and Zhang]{xie_atomistic_2011}
Xie,~Y.; Zhang,~J.~M. Atomistic simulation of phonon dispersion for
  body-centred cubic alkali metals. \emph{Canadian Journal of Physics}
  \textbf{2011}, \emph{86}, 801--805, DOI: \doi{10.1139/P07-200}\relax
\mciteBstWouldAddEndPuncttrue
\mciteSetBstMidEndSepPunct{\mcitedefaultmidpunct}
{\mcitedefaultendpunct}{\mcitedefaultseppunct}\relax
\EndOfBibitem
\bibitem[Balluffi \latin{et~al.}(2005)Balluffi, Allen, and
  Carter]{balluffi_kinetics_2005}
Balluffi,~R.~W.; Allen,~S.~M.; Carter,~W.~C. \emph{Kinetics of {Materials}},
  1st ed.; Wiley-Interscience: Hoboken, N.J, 2005\relax
\mciteBstWouldAddEndPuncttrue
\mciteSetBstMidEndSepPunct{\mcitedefaultmidpunct}
{\mcitedefaultendpunct}{\mcitedefaultseppunct}\relax
\EndOfBibitem
\bibitem[Xu \latin{et~al.}(2017)Xu, Ahmad, Aryanfar, Viswanathan, and
  Greer]{xu_enhanced_2017}
Xu,~C.; Ahmad,~Z.; Aryanfar,~A.; Viswanathan,~V.; Greer,~J.~R. Enhanced
  strength and temperature dependence of mechanical properties of {Li} at small
  scales and its implications for {Li} metal anodes. \emph{Proceedings of the
  National Academy of Sciences} \textbf{2017}, \emph{114}, 57--61, DOI:
  \doi{10.1073/pnas.1615733114}, Publisher: National Academy of Sciences
  Section: Physical Sciences\relax
\mciteBstWouldAddEndPuncttrue
\mciteSetBstMidEndSepPunct{\mcitedefaultmidpunct}
{\mcitedefaultendpunct}{\mcitedefaultseppunct}\relax
\EndOfBibitem
\bibitem[Owen and Williams(1954)Owen, and Williams]{owen_x-ray_1954}
Owen,~E.~A.; Williams,~G.~I. X-{Ray} {Measurements} on {Lithium} at {Low}
  {Temperatures}. \emph{Proceedings of the Physical Society. Section A}
  \textbf{1954}, \emph{67}, 895--900, DOI:
  \doi{10.1088/0370-1298/67/10/306}\relax
\mciteBstWouldAddEndPuncttrue
\mciteSetBstMidEndSepPunct{\mcitedefaultmidpunct}
{\mcitedefaultendpunct}{\mcitedefaultseppunct}\relax
\EndOfBibitem
\bibitem[C.~Nash and Smith(1959)C.~Nash, and Smith]{c_nash_single-crystal_1959}
C.~Nash,~H.; Smith,~C.~S. Single-crystal elastic constants of lithium.
  \emph{Journal of Physics and Chemistry of Solids} \textbf{1959}, \emph{9},
  113--118, DOI: \doi{10.1016/0022-3697(59)90201-X}\relax
\mciteBstWouldAddEndPuncttrue
\mciteSetBstMidEndSepPunct{\mcitedefaultmidpunct}
{\mcitedefaultendpunct}{\mcitedefaultseppunct}\relax
\EndOfBibitem
\bibitem[Trivisonno and Smith(1961)Trivisonno, and
  Smith]{trivisonno_elastic_1961}
Trivisonno,~J.; Smith,~C.~S. Elastic constants of lithium-magnesium alloys.
  \emph{Acta Metall.} \textbf{1961}, \emph{9}, 1064--1071\relax
\mciteBstWouldAddEndPuncttrue
\mciteSetBstMidEndSepPunct{\mcitedefaultmidpunct}
{\mcitedefaultendpunct}{\mcitedefaultseppunct}\relax
\EndOfBibitem
\bibitem[Slotwinski and Trivisonno(1969)Slotwinski, and
  Trivisonno]{slotwinski_temperature_1969}
Slotwinski,~T.; Trivisonno,~J. Temperature dependence of the elastic constants
  of single crystal lithium. \emph{Journal of Physics and Chemistry of Solids}
  \textbf{1969}, \emph{30}, 1276--1278, DOI:
  \doi{10.1016/0022-3697(69)90386-2}\relax
\mciteBstWouldAddEndPuncttrue
\mciteSetBstMidEndSepPunct{\mcitedefaultmidpunct}
{\mcitedefaultendpunct}{\mcitedefaultseppunct}\relax
\EndOfBibitem
\bibitem[Masias \latin{et~al.}(2019)Masias, Felten, Garcia-Mendez, Wolfenstine,
  and Sakamoto]{masias_elastic_2019}
Masias,~A.; Felten,~N.; Garcia-Mendez,~R.; Wolfenstine,~J.; Sakamoto,~J.
  Elastic, plastic, and creep mechanical properties of lithium metal.
  \emph{Journal of Materials Science} \textbf{2019}, \emph{54}, 2585--2600,
  DOI: \doi{10.1007/s10853-018-2971-3}\relax
\mciteBstWouldAddEndPuncttrue
\mciteSetBstMidEndSepPunct{\mcitedefaultmidpunct}
{\mcitedefaultendpunct}{\mcitedefaultseppunct}\relax
\EndOfBibitem
\bibitem[Wang \latin{et~al.}(2019)Wang, Dang, Wang, Xiao, and
  Cheng]{wang_mechanical_2019}
Wang,~Y.; Dang,~D.; Wang,~M.; Xiao,~X.; Cheng,~Y.-T. Mechanical behavior of
  electroplated mossy lithium at room temperature studied by flat punch
  indentation. \emph{Applied Physics Letters} \textbf{2019}, \emph{115},
  043903, DOI: \doi{10.1063/1.5111150}, Publisher: American Institute of
  Physics\relax
\mciteBstWouldAddEndPuncttrue
\mciteSetBstMidEndSepPunct{\mcitedefaultmidpunct}
{\mcitedefaultendpunct}{\mcitedefaultseppunct}\relax
\EndOfBibitem
\bibitem[Zhang \latin{et~al.}(2020)Zhang, Li, Djemia, Yang, and
  Hu]{zhang_prediction_2020}
Zhang,~H.; Li,~C.; Djemia,~P.; Yang,~R.; Hu,~Q. Prediction on temperature
  dependent elastic constants of “soft” metal {Al} by {AIMD} and {QHA}.
  \emph{Journal of Materials Science \& Technology} \textbf{2020}, \emph{45},
  92--97, DOI: \doi{10.1016/j.jmst.2019.11.029}\relax
\mciteBstWouldAddEndPuncttrue
\mciteSetBstMidEndSepPunct{\mcitedefaultmidpunct}
{\mcitedefaultendpunct}{\mcitedefaultseppunct}\relax
\EndOfBibitem
\bibitem[Nye(1985)]{nye_physical_1985}
Nye,~J.~F. \emph{Physical {Properties} of {Crystals}: {Their} {Representation}
  by {Tensors} and {Matrices}}; Clarendon Press, 1985\relax
\mciteBstWouldAddEndPuncttrue
\mciteSetBstMidEndSepPunct{\mcitedefaultmidpunct}
{\mcitedefaultendpunct}{\mcitedefaultseppunct}\relax
\EndOfBibitem
\bibitem[Jain \latin{et~al.}(2013)Jain, Ong, Hautier, Chen, Richards, Dacek,
  Cholia, Gunter, Skinner, Ceder, and Persson]{jain_commentary_2013}
Jain,~A.; Ong,~S.~P.; Hautier,~G.; Chen,~W.; Richards,~W.~D.; Dacek,~S.;
  Cholia,~S.; Gunter,~D.; Skinner,~D.; Ceder,~G.; Persson,~K.~A. Commentary:
  {The} {Materials} {Project}: {A} materials genome approach to accelerating
  materials innovation. \emph{APL Materials} \textbf{2013}, \emph{1}, 011002,
  DOI: \doi{10.1063/1.4812323}, Publisher: American Institute of Physics\relax
\mciteBstWouldAddEndPuncttrue
\mciteSetBstMidEndSepPunct{\mcitedefaultmidpunct}
{\mcitedefaultendpunct}{\mcitedefaultseppunct}\relax
\EndOfBibitem
\bibitem[Jäckle and Groß(2014)Jäckle, and Groß]{jackle_microscopic_2014}
Jäckle,~M.; Groß,~A. Microscopic properties of lithium, sodium, and magnesium
  battery anode materials related to possible dendrite growth. \emph{The
  Journal of Chemical Physics} \textbf{2014}, \emph{141}, 174710, DOI:
  \doi{10.1063/1.4901055}, Publisher: American Institute of Physics\relax
\mciteBstWouldAddEndPuncttrue
\mciteSetBstMidEndSepPunct{\mcitedefaultmidpunct}
{\mcitedefaultendpunct}{\mcitedefaultseppunct}\relax
\EndOfBibitem
\bibitem[Pande and Viswanathan(2019)Pande, and
  Viswanathan]{pande_computational_2019}
Pande,~V.; Viswanathan,~V. Computational {Screening} of {Current} {Collectors}
  for {Enabling} {Anode}-{Free} {Lithium} {Metal} {Batteries}. \emph{ACS Energy
  Letters} \textbf{2019}, \emph{4}, 2952--2959, DOI:
  \doi{10.1021/acsenergylett.9b02306}, Publisher: American Chemical
  Society\relax
\mciteBstWouldAddEndPuncttrue
\mciteSetBstMidEndSepPunct{\mcitedefaultmidpunct}
{\mcitedefaultendpunct}{\mcitedefaultseppunct}\relax
\EndOfBibitem
\bibitem[Sun and Ceder(2013)Sun, and Ceder]{sun_efficient_2013}
Sun,~W.; Ceder,~G. Efficient creation and convergence of surface slabs.
  \emph{Surface Science} \textbf{2013}, \emph{617}, 53--59, DOI:
  \doi{10.1016/j.susc.2013.05.016}\relax
\mciteBstWouldAddEndPuncttrue
\mciteSetBstMidEndSepPunct{\mcitedefaultmidpunct}
{\mcitedefaultendpunct}{\mcitedefaultseppunct}\relax
\EndOfBibitem
\bibitem[Henkelman \latin{et~al.}(2000)Henkelman, Uberuaga, and
  Jónsson]{henkelman_climbing_2000}
Henkelman,~G.; Uberuaga,~B.~P.; Jónsson,~H. A climbing image nudged elastic
  band method for finding saddle points and minimum energy paths. \emph{The
  Journal of Chemical Physics} \textbf{2000}, \emph{113}, 9901--9904, DOI:
  \doi{10.1063/1.1329672}, Publisher: American Institute of Physics\relax
\mciteBstWouldAddEndPuncttrue
\mciteSetBstMidEndSepPunct{\mcitedefaultmidpunct}
{\mcitedefaultendpunct}{\mcitedefaultseppunct}\relax
\EndOfBibitem
\bibitem[Bell and Hinshelwood(1997)Bell, and Hinshelwood]{bell_theory_1997}
Bell,~R.~P.; Hinshelwood,~C.~N. The theory of reactions involving proton
  transfers. \emph{Proceedings of the Royal Society of London. Series A -
  Mathematical and Physical Sciences} \textbf{1997}, \emph{154}, 414--429, DOI:
  \doi{10.1098/rspa.1936.0060}, Publisher: Royal Society\relax
\mciteBstWouldAddEndPuncttrue
\mciteSetBstMidEndSepPunct{\mcitedefaultmidpunct}
{\mcitedefaultendpunct}{\mcitedefaultseppunct}\relax
\EndOfBibitem
\bibitem[Evans and Polanyi(1936)Evans, and Polanyi]{evans_further_1936}
Evans,~M.~G.; Polanyi,~M. Further considerations on the thermodynamics of
  chemical equilibria and reaction rates. \emph{Transactions of the Faraday
  Society} \textbf{1936}, \emph{32}, 1333--1360, DOI:
  \doi{10.1039/TF9363201333}, Publisher: The Royal Society of Chemistry\relax
\mciteBstWouldAddEndPuncttrue
\mciteSetBstMidEndSepPunct{\mcitedefaultmidpunct}
{\mcitedefaultendpunct}{\mcitedefaultseppunct}\relax
\EndOfBibitem
\bibitem[Schwoebel and Shipsey(1966)Schwoebel, and
  Shipsey]{schwoebel_step_1966}
Schwoebel,~R.~L.; Shipsey,~E.~J. Step {Motion} on {Crystal} {Surfaces}.
  \emph{Journal of Applied Physics} \textbf{1966}, \emph{37}, 3682--3686, DOI:
  \doi{10.1063/1.1707904}, Publisher: American Institute of Physics\relax
\mciteBstWouldAddEndPuncttrue
\mciteSetBstMidEndSepPunct{\mcitedefaultmidpunct}
{\mcitedefaultendpunct}{\mcitedefaultseppunct}\relax
\EndOfBibitem
\bibitem[Biewald(2020)]{wandb}
Biewald,~L. Experiment Tracking with Weights and Biases. 2020;
  \url{https://www.wandb.com/}, Software available from wandb.com\relax
\mciteBstWouldAddEndPuncttrue
\mciteSetBstMidEndSepPunct{\mcitedefaultmidpunct}
{\mcitedefaultendpunct}{\mcitedefaultseppunct}\relax
\EndOfBibitem
\bibitem[Giannozzi \latin{et~al.}(2009)Giannozzi, Baroni, Bonini, Calandra,
  Car, Cavazzoni, Ceresoli, Chiarotti, Cococcioni, Dabo, Corso, Gironcoli,
  Fabris, Fratesi, Gebauer, Gerstmann, Gougoussis, Kokalj, Lazzeri,
  Martin-Samos, Marzari, Mauri, Mazzarello, Paolini, Pasquarello, Paulatto,
  Sbraccia, Scandolo, Sclauzero, Seitsonen, Smogunov, Umari, and
  Wentzcovitch]{giannozzi_quantum_2009}
Giannozzi,~P. \latin{et~al.}  {QUANTUM} {ESPRESSO}: a modular and open-source
  software project for quantum simulations of materials. \emph{Journal of
  Physics: Condensed Matter} \textbf{2009}, \emph{21}, 395502, DOI:
  \doi{10.1088/0953-8984/21/39/395502}\relax
\mciteBstWouldAddEndPuncttrue
\mciteSetBstMidEndSepPunct{\mcitedefaultmidpunct}
{\mcitedefaultendpunct}{\mcitedefaultseppunct}\relax
\EndOfBibitem
\bibitem[Perdew \latin{et~al.}(1997)Perdew, Burke, and
  Ernzerhof]{perdew_generalized_1997}
Perdew,~J.~P.; Burke,~K.; Ernzerhof,~M. Generalized {Gradient} {Approximation}
  {Made} {Simple} [{Phys}. {Rev}. {Lett}. 77, 3865 (1996)]. \emph{Physical
  Review Letters} \textbf{1997}, \emph{78}, 1396--1396, DOI:
  \doi{10.1103/PhysRevLett.78.1396}, Publisher: American Physical Society\relax
\mciteBstWouldAddEndPuncttrue
\mciteSetBstMidEndSepPunct{\mcitedefaultmidpunct}
{\mcitedefaultendpunct}{\mcitedefaultseppunct}\relax
\EndOfBibitem
\bibitem[Blöchl(1994)]{blochl_projector_1994}
Blöchl,~P.~E. Projector augmented-wave method. \emph{Physical Review B}
  \textbf{1994}, \emph{50}, 17953--17979, DOI: \doi{10.1103/PhysRevB.50.17953},
  Publisher: American Physical Society\relax
\mciteBstWouldAddEndPuncttrue
\mciteSetBstMidEndSepPunct{\mcitedefaultmidpunct}
{\mcitedefaultendpunct}{\mcitedefaultseppunct}\relax
\EndOfBibitem
\bibitem[Monkhorst and Pack(1976)Monkhorst, and Pack]{monkhorst_special_1976}
Monkhorst,~H.~J.; Pack,~J.~D. Special points for {Brillouin}-zone integrations.
  \emph{Physical Review B} \textbf{1976}, \emph{13}, 5188--5192, DOI:
  \doi{10.1103/PhysRevB.13.5188}\relax
\mciteBstWouldAddEndPuncttrue
\mciteSetBstMidEndSepPunct{\mcitedefaultmidpunct}
{\mcitedefaultendpunct}{\mcitedefaultseppunct}\relax
\EndOfBibitem
\bibitem[Methfessel and Paxton(1989)Methfessel, and
  Paxton]{methfessel_high-precision_1989}
Methfessel,~M.; Paxton,~A.~T. High-precision sampling for {Brillouin}-zone
  integration in metals. \emph{Physical Review B} \textbf{1989}, \emph{40},
  3616--3621, DOI: \doi{10.1103/PhysRevB.40.3616}, Publisher: American Physical
  Society\relax
\mciteBstWouldAddEndPuncttrue
\mciteSetBstMidEndSepPunct{\mcitedefaultmidpunct}
{\mcitedefaultendpunct}{\mcitedefaultseppunct}\relax
\EndOfBibitem
\bibitem[Zhang \latin{et~al.}(2020)Zhang, Wang, Chen, Zeng, Zhang, Wang, and
  E]{zhang_dp-gen_2020}
Zhang,~Y.; Wang,~H.; Chen,~W.; Zeng,~J.; Zhang,~L.; Wang,~H.; E,~W. {DP}-{GEN}:
  {A} concurrent learning platform for the generation of reliable deep learning
  based potential energy models. \emph{Computer Physics Communications}
  \textbf{2020}, 107206, DOI: \doi{10.1016/j.cpc.2020.107206}, arXiv:
  1910.12690\relax
\mciteBstWouldAddEndPuncttrue
\mciteSetBstMidEndSepPunct{\mcitedefaultmidpunct}
{\mcitedefaultendpunct}{\mcitedefaultseppunct}\relax
\EndOfBibitem
\end{mcitethebibliography}

\section*{Acknowledgments}
V.V. and M.K.P. acknowledge support from Google Collabs. This material is based on work that is partially funded by Google. A.M. was supported by the U.S. Department of Energy, Office of Science, Office of Advanced Scientific Computing Research, Department of Energy Computational Science Graduate Fellowship under Award Number DE-SC0021110. M.K.P., A.M.Y. and V.V. also acknowledge Extreme Science and Engineering Discovery Environment (XSEDE) for providing computational resources through Award No. TG-CTS180061.\\
 
\section*{Author Contributions}
Conceptualization, M.K.P, V.V. and E.K.D; Methodology - Modeling - M.K.P., S.B., A.M.; Investigation, M.K.P.; Formal Analysis, M.K.P. and V.V.; Data Curation, M.K.P. and A.M.Y; Writing - Original Draft, M.K.P. and V.V.; Writing - Review \& Editing - M.K.P. and V.V.; Supervision. E.D.C., B.K. and V.V. \newline

\section*{Competing Interests}
There are no competing interests to the authors' knowledge.
\newline

\section*{Data and Materials Availability}
 All code and supporting data used in the work will be made available on Zenodo and OpenKIM database.
 \newline 
\section*{Supplementary materials}
\vspace{-0.5cm}
\hspace{2pt}\\
Supplementary Text\\
Figs. S1\\

\newpage

\clearpage

\end{document}


\subsection*{Methodology of Calculations in Table 1}
\subsubsection{Monovacancy Formation Energy(E$_v$)}
$E_v$ is calculated using the formula
$$ E_v=E_{\mathrm{bulk\ with\ vacancy}}-\frac{N-1}{N}E_{\mathrm{bulk\ without\ vacancy}} $$
where N is the number of atoms in the bulk structure without any vacancies. The reported value is that for a cubic BCC supercell of size 5x5x5, the maximum size we calculated with DFT.
\\
\\
\subsubsection{Elastic Constants}
The elastic constants were calculated using the same method as Ahmad et al. (29) as an expansion of the energy about the equilibrium position with different volume conserving strains. The free energy $F$ can be expanded as a function of the strains $e_i$ in Voigt notation as 
\begin{equation}
    F = F_0 + \frac{V}{2}\sum\limits^6_{i=1}\sum\limits^6_{j=1}C_{ij}e_ie_j + O(e_i^3)
\end{equation}

where $F_0$ is an arbitrary constant and $C_ij$ are the components of the elastic tensor. In practice since we are only predicting the 0K elastic constants, the DFT energy is equivalent to the Free Energy. Combinations of the elastic constants can therefore be extracted as in any standard textbook. In our work we use the bulk modulus 
\begin{equation}\label{eq:k}
    K = (C_{11}+2C_{12})/3
\end{equation} which is obtained by fitting the Birch-Murnaghan equation of state.
Additionally, the volume-conserving orthorhombic strain with $e_1=-e_2=x$, $e_3=x^2/(4-x^2)$ and $e_4=e_5=e_6=0$ to get 
\begin{equation}\label{eq:cprime}
    C'=C_{11} - C_{12}.
\end{equation}
Eq. \ref{eq:k} and eq. \ref{eq:cprime} can be solved to isolate $C_{11}$ and $C_{12}$. 
Finally, the volume-conserving monoclinic strain with $e_4=x$, $e_1=x^2/(4-x^2)$ and $e_2=e_3=e_5=e_6=0$ is used to find $C_{44}$.

\section*{Surface Potential Energy Surfaces}

\begin{figure}[H]
\centering
\includegraphics[height=1.2\textwidth]{SI/SI_NequIP64_pes.png}
\caption{Surface Potential Energy Surfaces for Miller indices up to 3 for BCC lithium}
\label{fig:surface_pes_all}
\end{figure}
